\documentclass[preprint2]{aastex}

\newcommand{\oiii}{[\ion{O}{3}]}
\newcommand{\oii}{[\ion{O}{2}]}
\newcommand{\oi}{[\ion{O}{1}]}
\newcommand{\nii}{[\ion{N}{2}]}
\newcommand{\neiii}{[\ion{Ne}{3}]}
\newcommand{\ariii}{[\ion{Ar}{3}]}
\newcommand{\ariv}{[\ion{Ar}{4}]}
\newcommand{\sii}{[\ion{S}{2}]}
\newcommand{\siii}{[\ion{S}{3}]}
\newcommand{\fevi}{[\ion{Fe}{6}]}
\newcommand{\cliii}{[\ion{Cl}{3}]}
\newcommand{\heii}{He\,{\sc ii}}
\newcommand{\hei}{He\,{\sc i}}
\newcommand{\rniii}{\ion{N}{3}}
\newcommand{\rciii}{\ion{C}{3}}
\newcommand{\roii}{\ion{O}{2}}
\newcommand{\moiii}{m($5007$)}
\newcommand{\mabsoiii}{M($5007$)}
\newcommand{\kms}{km~s$^{-1}$}
\newcommand{\msun}{M$_\odot$}
\newcommand{\lsun}{L$_\odot$}
\newcommand{\Ne}{$N_e$}
\newcommand{\Te}{$T_e$}
\newcommand{\avg}[1]{\left< #1 \right>}
\newcommand{\ph}{\phantom}
\newcommand{\tnm}{\tablenotemark}
\newcommand{\ta}{\tablenotemark{a}}
\newcommand{\tb}{\tablenotemark{b}}
\newcommand{\mc}{\multicolumn}
\newcommand{\nd}{\nodata}

\shorttitle{PNe in the outskirts of M31}
\shortauthors{Corradi et al.}

\begin{document}


\title{The chemistry of planetary nebulae in the outer regions of M31}


\author{
R.L.M. Corradi\altaffilmark{1,2}, 
K.B. Kwitter\altaffilmark{3}, 
B. Balick\altaffilmark{4},
R.B.C. Henry\altaffilmark{5}, and
K. Hensley \altaffilmark{3}}


\altaffiltext{1}{Instituto de Astrof{\'{\i}}sica de Canarias, E-38200
  La Laguna, Tenerife, Spain}
\altaffiltext{2}{Departamento de Astrof{\'{\i}}sica, Universidad de La
  Laguna, E-38206 La Laguna, Tenerife, Spain}
\altaffiltext{3}{Department of Astronomy, Williams College, Williamstown, MA 01267, USA}
\altaffiltext{4}{Department of Astronomy, University of Washington, Seattle, WA 98195-1581, USA}
\altaffiltext{5}{H.L. Dodge Department of Physics \& Astronomy, University of Oklahoma, Norman, OK 73019, USA}


\begin{abstract}
We present spectroscopy of nine planetary nebulae (PNe) in the
outskirts of M31, all but one obtained with the 10.4~m GTC
telescope.  These sources extend our previous study of the oxygen
abundance gradient of M31 to galactocentric radii as large as
100~kpc. None of the targets are bona fide members of a classical,
metal-poor and ancient halo. Two of the outermost PNe have solar oxygen
abundances, as well as radial velocities consistent with the
kinematics of the extended disk of M31.  The other PNe have a slightly
lower oxygen content ([O/H]$\sim$$-0.4$) and in some cases large
deviations from the disk kinematics.  These PNe support the current
view that the external regions of M31 are the result of a complex
interaction and merger process, with evidence for a widespread
population of solar--metallicity stars produced in a starburst that 
occurred $\sim$2~Gyr ago.
\end{abstract}



\section{Introduction}

Planetary nebulae (PNe) are valuable tracers of stellar populations in
all types of galaxies. They are widely used as distance indicators, via
the invariant bright cutoff of the PN luminosity function 
\citep[PNLF; e.g.][]{c02}, and as tracers of the luminosity \citep{bu06},
dynamics \citep[e.g.][]{m06} and chemistry (e.g. \citealt{k12,b13},
hereafter papers I and II, respectively) of their stellar progenitors.

Recently, PNe have been used to determine the metallicity gradient in
the disk of the nearby spiral galaxies M33 \citep{m09} and M31 (papers
I and II).  These gradients provide direct constraints to models of
disk formation and evolution. In particular, measurements in the outer
regions of the disks have the power to test novel ideas such as the
importance of external disturbances to produce extended young stellar
disks \citep{w11}. In the case of M31, it has been proposed that a
tidal encounter with M33 that occurred about 3 Gyr ago would have
produced a vast extended disk with homogeneous metallicity
\citep{b12,b15}.  Such an extended disk has been reported before
\citep{ib05}, and PNe provide the opportunity to test its predicted
chemical content. In earlier papers, we have studied PNe out to
deprojected galactocentric distances of 60~kpc, finding nearly solar
O/H abundances and a flat O/H gradient in these regions.  In this
work, we extend our study by obtaining high-quality spectra of another
nine PNe at larger radii in the outskirts of M31.

\begin{deluxetable}{lcccrrrrrcc}
\tabletypesize{\footnotesize}
\tablecolumns{10}
\tablewidth{0pc}
\tablecaption{Basic properties of the target PNe\label{T-target}}
\tablehead{
Name & \mc{2}{c}{R.A.\tnm{a}\,\ph{pipppp}Dec\tnm{a}} & \moiii\tnm{a} & $\xi$\ph{p} & $\eta$\ph{p} & $d_{app}$ & $d_{app}$ &$d_{disk}$ & V$_{\odot,sys}$\tnm{a} &V$_{diff}$ \\
     &  [J2000] & [J2000]                    &   [mag]         & [deg]       & [deg]        &  [deg]   &  [kpc]   & [kpc]    & [\kms] & [\kms]}
\startdata
M2507                    & 00 48 27.2 & +39 55 34.3 & 21.23     &  1.10 & -1.33 & 1.73 &  23.2 &  106.3 & -147      & 180     \\	
M2538                    & 00 36 28.8 & +39 35 26.4 & 20.25     & -1.21 & -1.67 & 2.06 &  27.7 &   28.0 & -426      & 110     \\	
M2539                    & 00 36 12.6 & +39 35 41.9 & 21.16     & -1.26 & -1.66 & 2.08 &  28.0 &   28.0 & -426      & 110     \\	
M2541                    & 00 35 09.1 & +39 28 25.2 & 21.78     & -1.46 & -1.78 & 2.30 &  31.0 &   31.3 & -456      & \ph{0}80\\	
M2543                    & 00 35 50.7 & +42 21 04.5 & 21.61     & -1.27 &  1.09 & 1.68 &  22.6 &  105.8 & -272      & \ph{0}40\\	
M2549                    & 00 36 27.2 & +42 06 21.9 & 21.35     & -1.17 &  0.85 & 1.44 &  19.4 &   90.9 & \ph{0}-17 & 290     \\	
M2566                    & 00 49 28.3 & +40 59 53.9 & 21.26     &  1.27 & -0.26 & 1.30 &  17.4 &   73.8 & -247      & \ph{0}10\\	
M2988                    & 00 52 00.0 & +43 03 23.5 & 21.98     &  1.69 &  1.81 & 2.48 &  33.3 &   36.2 & \ph{0}-98 & \ph{0}20\\
M31-372\tablenotemark{b} & 00 46 41.5 & +43 59 03.7 & 22.6\ph{0}&  0.71 &  2.72 & 2.81 &  37.8 &   77.6 & \ph{0}-60 & 100     \\ 	
\enddata
\tablenotetext{a}{From \cite{m06}, except for M31-372.}
\tablenotetext{b}{Coordinates are from \cite{jf86}. V$_{sys}$ (uncertainty 40~\kms) and \moiii\ (uncertainty 0.3~mag) are estimated from our spectrum.}
\tablecomments{The following parameters for M31
have been assumed throughout the paper: distance 770~kpc \citep{fm90}; 
center at RA(J2000)= 0 42 44.3 and Dec(J2000)=+41 16 09.0 \citep{m06};
disk inclination $i$=77\degr.7 \citep{dv58} and position angle PA=37\degr.7 \citep{m06}; 
heliocentric systemic velocity V$_{sys}$=$-309$~kms\ \citep{m06}.}
\end{deluxetable}

\begin{figure*}[!ht]
\epsscale{2.0}
\plotone{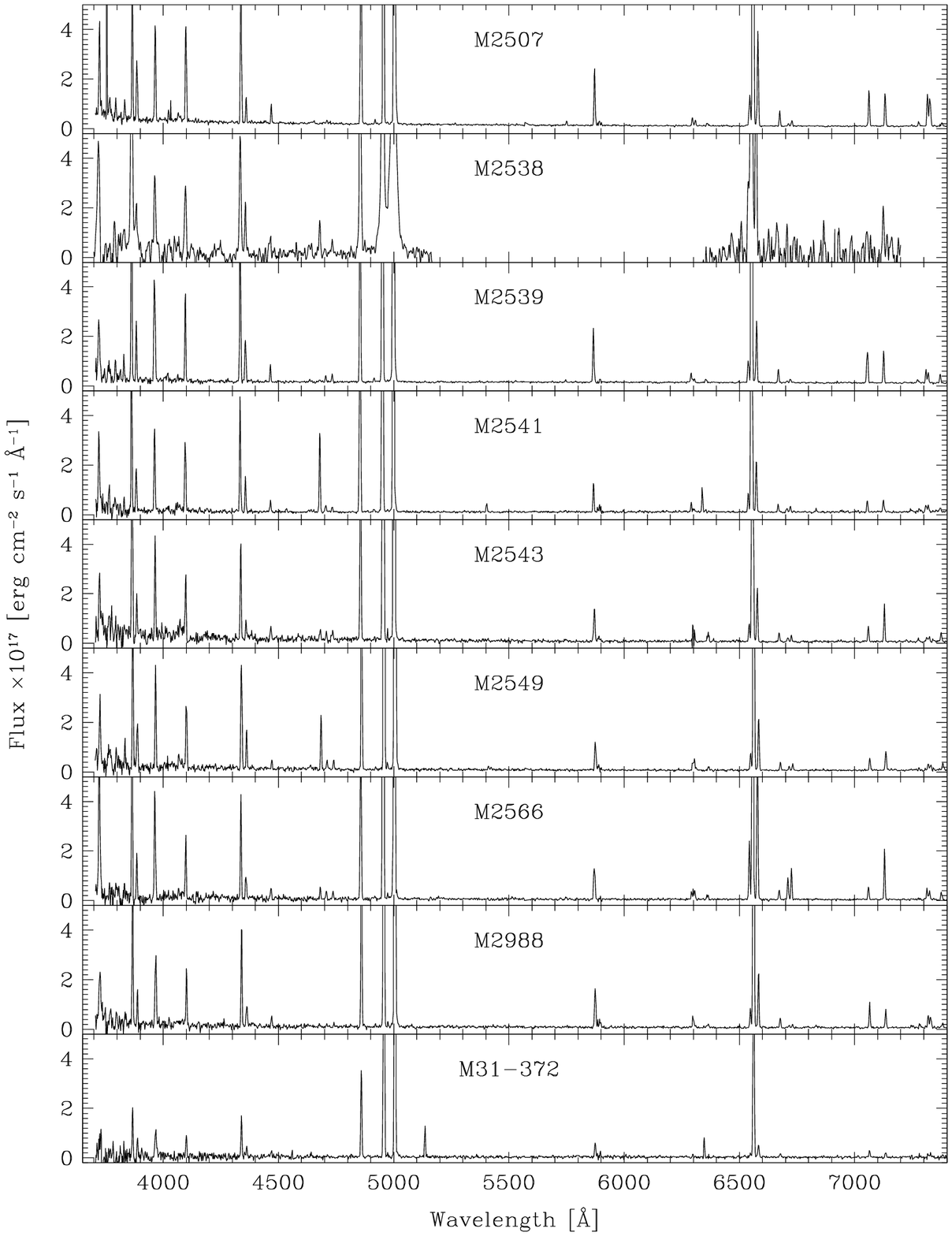}
\caption{The spectra of the target PNe. For the lower S/N spectrum of
  M2538, regions with high noise have been masked, and the whole
  spectrum was smoothed with a boxcar of three pixels.}
\label{F-spectra}
\end{figure*}

\section{Observations}

The target PNe are listed in Table~\ref{T-target}. All but one were
selected from the list of \citet{m06} to be at large galactocentric
distances and sufficiently bright to obtain high quality spectra with
a 10m telescope. The putative halo PN M31-372 observed by \citet{jf86}
was additionally included.

Spectroscopy of eight of these PNe was obtained in different nights in
October 2013 in service-queue mode at the 10.4m~GTC telescope on the
island of La Palma, Spain. The OSIRIS instrument was used in its
longslit mode. The combination of grism R1000B and a slit width of
$0''.8$ provides a spectral dispersion of 0.21~nm per (binned
$\times$2) pixel, a resolution of 0.63~nm, and a spectral coverage
from 370 to 785~nm. Seeing varied between $0''.6$ and $1''.1$ (full
width at half maximum), and spatial scale along the slit was $0''.254$
per binned pixel. Data were generally obtained under photometric
weather conditions and grey moon. The slit was oriented along the
  parallactic angle for all targets and standard stars. Total
exposure times per target varied from 120 to 135 min, depending on the
brightness of the source, split into three or four sub-exposures. The
GTC calibration plan provides, in each observing night, at least one
spectrophotometric standard to be used for flux calibration.

Additionally, the PN M2538 was observed with the Dual-Imaging
Spectrograph at Apache Point Observatory (New Mexico, USA) in 2011
October. We observed through a 2$''$ slit oriented along the
  parallactic angle. The B1000 grating provided coverage from 370 to
505~nm at 0.25~nm resolution; the R300 grating covered 520--960~nm at
0.70~nm resolution.
Standard bias, flatfield, and
emission-lamp exposures were taken and applied to the data. Flux
calibration was obtained via observations of the standard star BD+28
4211. We obtained six 20~min integrations of M2538, which were
co-added after calibration.

As PNe at the distance of M31 (770~kpc, see Table~\ref{T-target})
are spatially unresolved, 1--D spectra were extracted and reduced
using {\it twodspec} in IRAF\footnote{IRAF is distributed by the
  National Optical Astronomy Observatory, which is operated by the
  Association of Universities for Research in Astronomy (AURA) under
  cooperative agreement with the National Science Foundation.}.

\section{Physics and chemistry of the nebulae}\label{S-abund}

The spectra of the target PNe are displayed in
Figure~\ref{F-spectra}. Emission-line fluxes were measured by
multi-Gaussian fit using {\it splot}. These fluxes are the input for
the abundance determinations, carried out with ELSA, our five-level
atom code \citep{j06}. We used the same method as in Papers I and II,
in order to produce a homogeneous set of data and abundance
estimations.  The reader is referred to those articles and to
  \citet{mi10} for the details of the analysis. The procedure includes
corrections of the observed line fluxes for interstellar reddening,
using the law of \citet{sm90}, and for the contamination of the
hydrogen Balmer lines by coincident recombination lines of He$^{++}$.
Dereddened fluxes were used to determine the electron temperature and
density \Ne\ and \Te\ with the appropriate line diagnostics. Note that
\Te(\oiii) was always properly determined, as the auroral line
\oiii4363 was accurately detected in all PNe
(Figure~\ref{F-spectra}). The same applies for \Ne(\sii), except for
M31-372 and M2538 where the \sii6717,6731 doublet could not be
measured and a default \Ne\ of 10000~cm$^{-3}$ was assumed.  In PNe
where no direct low-ionization temperature diagnostic has been
observed, the determination of the temperature to be used for
calculating low-ionization abundances (\Te(\nii)) depends on whether
\heii4686 is observed. If it is, we adopt the carefully derived
result from \citet{kaler86} that applies under this condition,
i.e. \Te(\nii)$=$10\,300~K.  If not, then we derive \Te(\nii) from
\Te(\oiii) according to the prescription from \citet{p92}.

The detected lines allowed the calculation of the abundances of
several O, Ar, N, S, and Ne ions. Total abundances were then
determined using ionization correction factors ({\it icf}) calculated
as described in \citet{kh01}.  The observed emission-line
measurements, logarithmic reddening parameter c(H$\beta$), \Te\ and
\Ne, ionic and total abundances are listed in the tables in
Appendix~A. 
They indicate that none of the PNe presented in this paper fulfill the
original definition by \citet{p83} for Galactic Type~I PNe of having
either $\log(N/O)>-0.3$ or He/H$>0.125$. This also applies to most of
PNe studied in papers I and II, indicating at most a very moderate N
and He enrichment for these bright PNe of M31. The PNe studied in
  this article also follow the relationships between O/H and other elemental
  abundances (such as N/O, Ne/H, and Ar/H) discussed in paper I.

\begin{figure}[!ht]
\epsscale{1.0}
\plotone{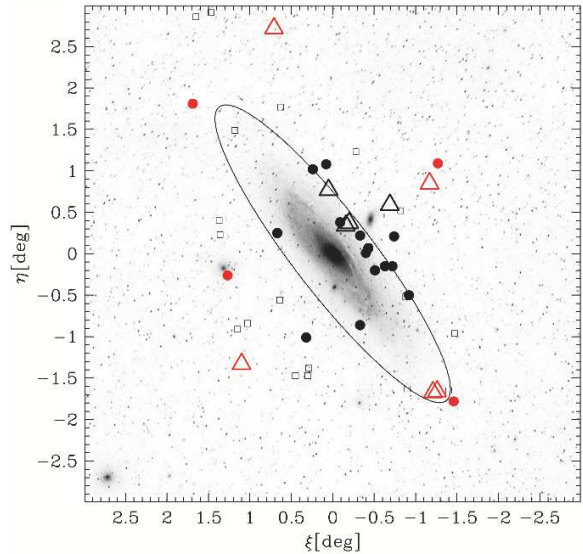} 
\caption{Location of the target PNe, overlaid on the DSS optical image. 
Circles and triangles
  indicate PNe with deviations from the model disk kinematics
  smaller and larger than 100~\kms, respectively (see text). Red symbols are the
  PNe presented in this work, and black symbols those discussed in
  papers I and II. Squares indicate the HST/ACS fields studied by
  \citet{b15}.  
The ellipse indicates the R$_{25}$  radius of M31.}
\label{F-location}
\end{figure}

\section{Location of targets within M31}\label{S-location}

Figure~\ref{F-location} shows the position in the sky of the target
PNe, including those discussed in papers I and II.  The geometric
transformations of \citet{h91}, with the parameters indicated in the
Note of Table~\ref{T-target}, were adopted to determine the R.A. and
Dec offsets $\xi$ and $\eta$ from the center of M31. These offsets,
and the total apparent distance $d_{app}$ in the plane of the sky, are
also listed in Table~\ref{T-target}.  The ellipse in the figure marks the
R$_{25}$ radius of M31 ($\sim$30~kpc, see paper II). It corresponds to
$\sim$5 scale-lengths of the exponential bright disk of M31
\citep{wk88}, showing that the newly observed PNe are located well
outside the familiar bright disk of M31.

These outer regions of M31 are surprisingly complex. A huge halo with
a radius $\ge$300~kpc and a smooth star density distribution has been
identified \citep{ib07,ib14}.  A number of structures, such as
streams, loops, and overdensity regions are seen in projection
throughout the halo \citep{l13,ib14}.  They are mostly associated with
accretion of satellite dwarf galaxies, as expected in the standard
hierarchical halo formation scenario. The most prominent structure is
the so-called Giant Stellar Stream (GSS), which is likely the latest,
most metal enhanced accretion event associated with the streams.  In
addition, outside the usual bright disk of M31, a smooth ``exodisk''
was found to extend out to a galactocentric distance of $\sim$40~kpc,
with detections as far as $\sim$70~kpc \citep{ib05}.  A set of
globular clusters in this zone forms a coherent kinematic system that
is similar to an extrapolation of M31's inner disk \citep{v14}.

As far as metallicity is concerned, there is a significant overlap
between the stellar [Fe/H] content of the different structures in the
outer regions of M31.  Overall, the mean halo metallicity decreases
with radius from [Fe/H]$\sim$$-0.7$ at $R\le$30~kpc to
[Fe/H]$\sim$$-1.5$ at $R\sim$150~kpc, but the smooth halo component
seems 0.2 dex more metal poor at all radii, and substructures span a
wide range in metallicity covering roughly two orders of
  magnitude \citep{ta10,ib14}. The strongest evidence of metal-rich
stars in the halo (beyond the bound satellite galaxies) at
$\avg{[\mbox{Fe/H}]}$$\sim$$-0.5$ is found in the GSS \citep{ib14}. As
with the extended exodisk,
\citet{b15} find an average value of [Fe/H]$\sim$$-0.3$ for the
several disk-like fields considered.  \citet{ch06} find several fields
in M31's exodisk in which AGB stars have roughly solar metal
abundances.  Figure 8 in \citep{ib14} illustrates the complexity of
the metallicity distribution in the outer regions of M31 caused by the
overlapping halo (with its ancient and more recent components) and the
extended disk.

Adopting the inclination and position angle of M31's inner disk, the
deprojected distances in the plane of the disk for the target PNe,
$d_{disk}$, are indicated in Table~\ref{T-target}.  The radial
velocities of PNe provide useful insights into their possible
relation to M31's extended disk. They are adopted from
\citet{m06}, except for M31--372 for which a rough estimate could be
obtained from our spectra (Table~\ref{T-target}). We compared them with
the average velocity expected at each position according to the
kinematic model of the extended disk presented by \citet{ib05}. The
most significant differences (V$_{diff}\ge100$~\kms, see
Table~\ref{T-target}) are found in PNe M2507, M2538, M2539, M2549, and
M31-372, suggesting that they may instead be associated with the halo
or some of its substructures.

\citet{b15} present colour-magnitude diagrams in fourteen HST/ACS
fields sampling galactocentric distances similar to the PNe presented
in this work. They cover regions of the extended disk of M31, or some
of the substructures in the halo, none representing the metal--poor
halo component.  The positions in the sky of the HST fields are
indicated by squares in Figure~\ref{F-location}. While some of these
fields are not far from our PNe, no noteworthy association can be
identified, except for the case of M2538, M2539, and M2541 which seem
related to the so-called Warp and G1 Clump, which are both disk
components (but note the abovementioned deviations from the disk
kinematics of two of these PNe).

\begin{figure}[!t]
\epsscale{1.0}
\plotone{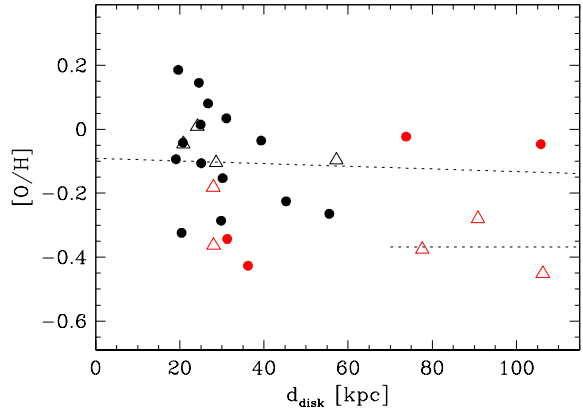}
\caption{The O/H abundance radial gradient.  Symbols are
  like in Figure~\ref{F-location}. Dashed lines are described in the text.}
\label{F-gradient}
\end{figure}

\section{M31 O/H abundance gradient}\label{S-grad}

The O/H abundance radial gradient, compiled with all PNe studied in
papers I and II (black symbols) and this work (red symbols), is
presented in Figure~\ref{F-gradient}. It extends the metallicity
gradient from this type of object to galactocentric distances larger
than 100~kpc, assuming that all PNe are located in the plane of the
disk of M31.  Sources with radial velocities consistent with the disk
kinematics (V$_{diff}$$<$100~\kms, circles)
are separated from those with large deviations from the disk
kinematical model of \citet[][triangles]{ib05}.  The graph shows that 
even at large distances PNe with solar metallicity and disk-like
kinematics exist. A least-square-fit to sources with V$_{diff}$$<$100~\kms\
provides an almost flat metallicity slope
([O/H]$=$$-0.0909$$-0.0004$$\times d_{disk}$, with
$\avg{[\mbox{O/H}]}$$=$$-0.11$). It is indicated by the upper dashed
line in the figure. As a note of caution, the two PNe with solar
oxygen abundance at $d_{disk}>60$~kpc are M2543 and M2566. Both are
located along the minor axis of M31's disk. Therefore their
deprojection factors to the disk plane are large, and radial
velocities are close to the systemic velocity of the galaxy. 
They do not appear to be associated with any major 
substructure in the halo \citep[see also][]{m06}, although it should also be noted that that
the GSS pervades a large portion of the outer regions of M31 out to
100 kpc, and its debris extend beyond the SE quadrant where it is more
prominent \citep{ib14}.

Figure~\ref{F-gradient} also shows that the three PNe at
$d_{disk}>60$~kpc which deviate from the disk kinematics have a lower
metallicity ($\avg{[O/H]}$$=$$-0.37$, lower dashed line). This
indicates that they might be related to an older population, but
  their only slightly subsolar O/H seemingly precludes their
  association with the stars in the most metal-poor and ancient
  component of the halo identified by e.g. \citet{ib14}.

\begin{figure}[!ht]
\epsscale{1.0}
\plotone{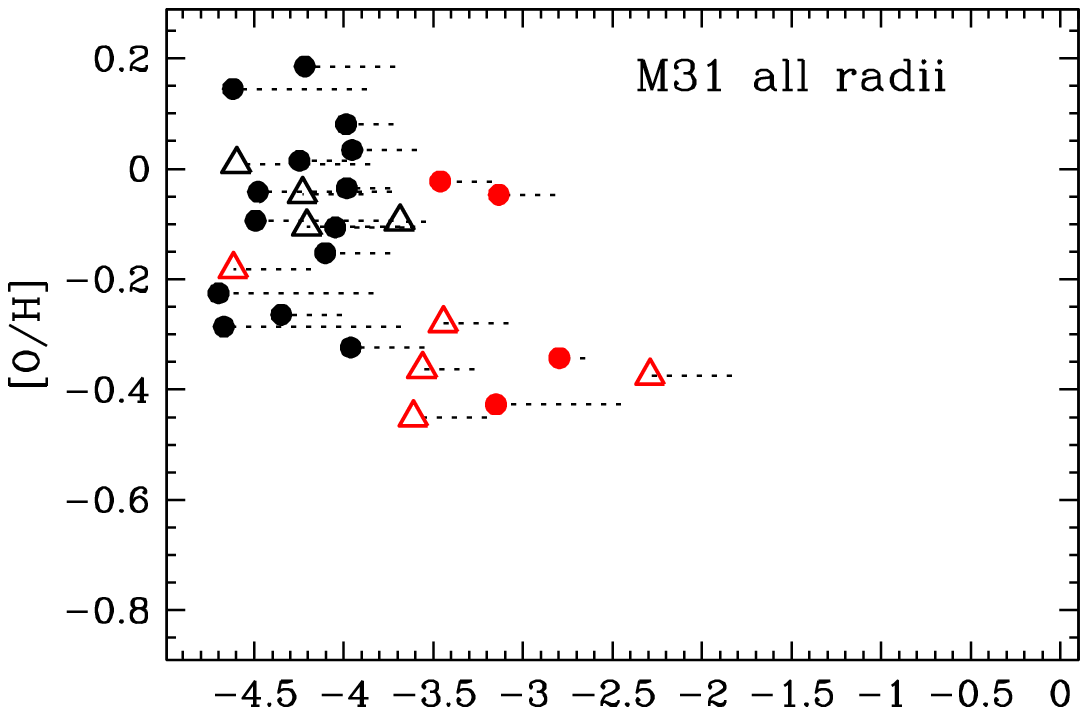}
\plotone{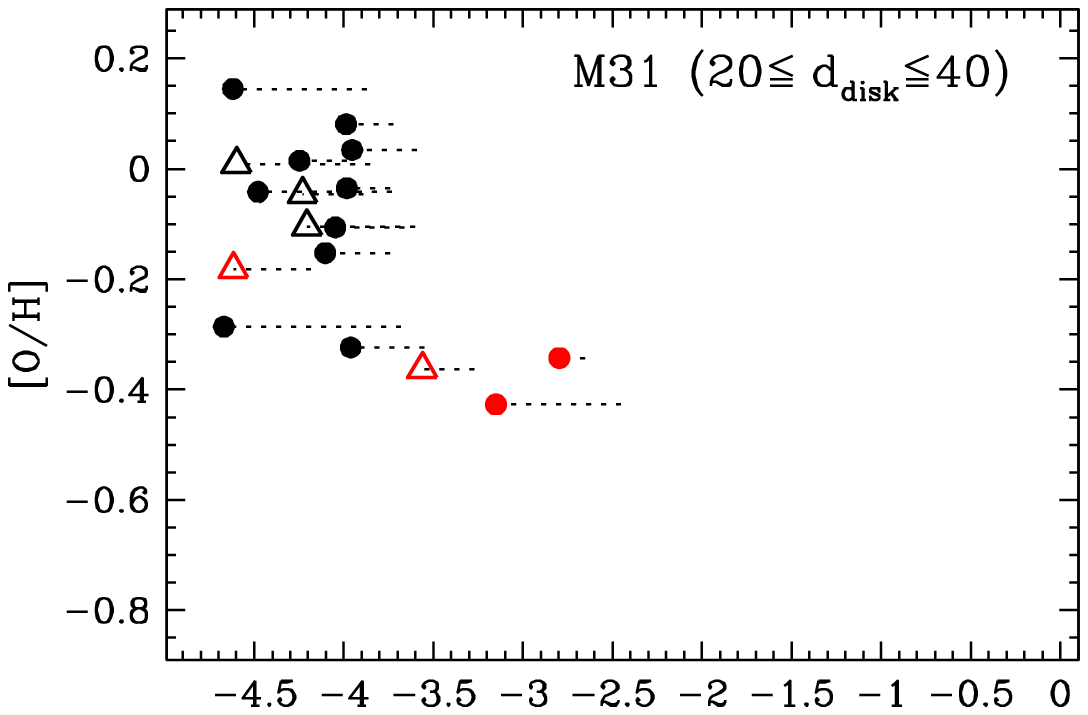}
\plotone{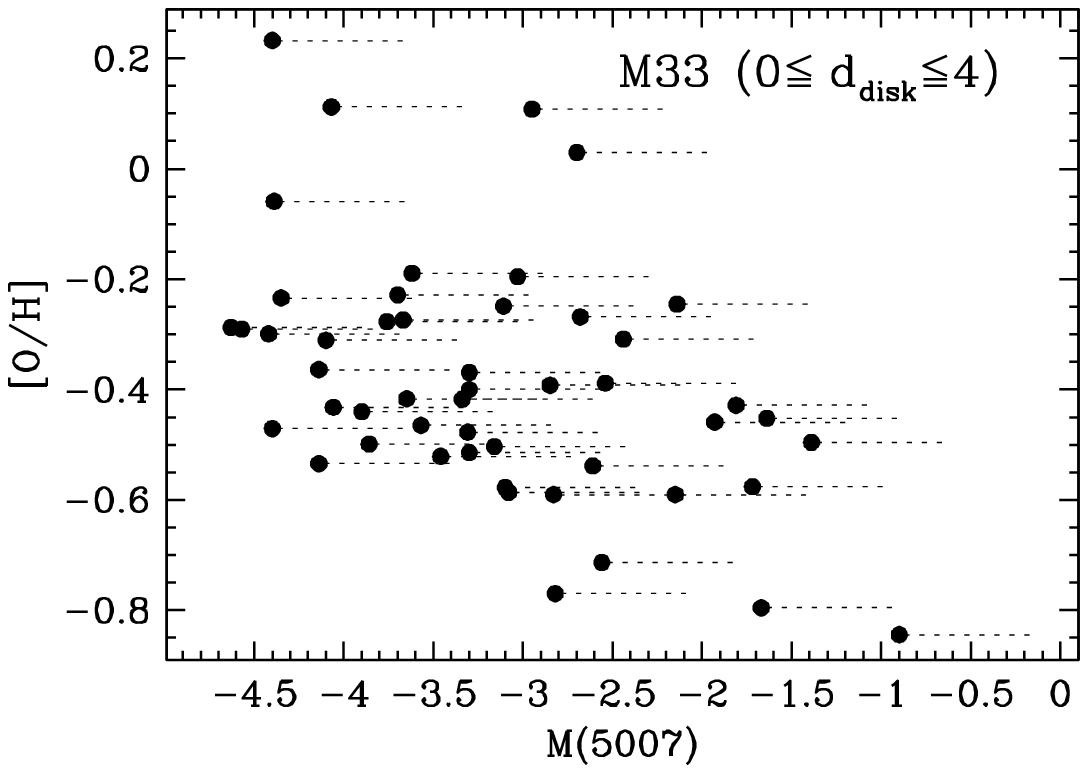}
\caption{The O/H abundances as a function of the 
  absolute \oiii\ magnitudes of PNe in M31 (top two panels, adopted
  distance 770~kpc) and in M33 (bottom panel, adopted distance
  840~kpc).  The dashed lines indicate the adopted extinction
  values (see text). For M31, symbols are  like in Figure~\ref{F-location}.}
\label{F-magabun}
\end{figure}

Before moving to a more general interpretation, two things should be
noted. In this paper, we assume that the O/H abundance ratio in the
PNe reflects the original oxygen content of the stellar progenitors,
i.e. of the ISM from which these stars formed.  There are some
indication that oxygen may be enhanced in some AGB stars during the
third dredge-up episodes, but this possible effect is neglected here
as no clear conclusion has been drawn yet \citep{k14,h12,di15}.  


Also, it should be recalled that our target PNe were selected to
  be bright enough to allow an accurate chemical analysis. Indeed, all
  lie within 2.4~mag from the bright cutoff of the PNLF.  \citet{r93}
  in the LMC, and \citet{jc99} in M31, suggested that there might be a
  slight dependence of metallicity with the \oiii\ luminosity, but
  concluded that the luminosities of the brightest PNe are nearly
  independent of O/H.  On the other hand, \citet{m04} did not find any
  trend of chemical abundances as a function of the \oiii\ luminosity
  for PNe in the LMC and our Galaxy, concluding that the chemical
  abundances derived from the brightest PNe are representative of the
  total PN population.  Figure~\ref{F-magabun}, compiled with our data
  of M31, shows some tendency of decreasing metallicity with
  \oiii\ luminosity. The upper panel displays our measured O/H
  abundances as a function of the absolute \oiii\ magnitude corrected
  for foreground and internal extinction using the c(H$\beta$) values
  determined in our spectra.  PNe in the brightest magnitude bin
  (\mabsoiii$\le$$-3.7$) have some spread in metallicity, but on
  average have higher O/H abundances than PNe in the next magnitude
  bin. To minimize the effects of the -- albeit shallow -- abundance
  gradient, the middle panel confirms that the trend persists if
  sources are selected in a more limited range of galactocentric
  distances.  A similar relation is also seen seen in M33 (lower
  panel), where we have used the data from \citet{m09}, which were
  dereddened adopting the M33 foreground and internal extinction from
  Cepheids \citep{f01}. The decrease of mean metallicity with the
\oiii\ luminosity may explain why the dispersion in the radial
gradient of M31 at $d_{disk}<40$~kpc is increased, compared to the
corresponding graphs in papers I and II, with the addition of the
slightly fainter PNe presented in this work.

In a standard scenario where metallicity of a galaxy increases
  with time and the PN progenitors do not form or destroy oxygen, the
  observed correlation between luminosity and O/H content is
  qualitatively consistent with the hypothesis that the most luminous
  PNe are produced by younger, i.e. more massive progenitors than
  fainter PNe. The hydrodynamical simulations by \citet[][see also
    \citealt{men08} and \citealt{c10}]{sch07} support such a view and
  provide further constraints to the stellar masses of our target
  PNe. \citet{sch07} found that only PNe with central stars masses
  $>$0.6~\msun\ can attain the \oiii\ luminosity of the bright cutoff
  of the PNLF if accompanied by sufficiently delayed optically
  thin/thick transition of the nebular gas.  Adopting the empirical
  initial-to-final mass relationships of \citet{cat08}, these
  relatively high core masses imply progenitors with an initial mass
  larger than $\sim$2~\msun.  Our new targets are between 1 and 2
  magnitudes below the PNLF cutoff. These slightly fainter magnitudes
  can either correspond to the most luminous PNe that have started to
  fade, or by PNe from slightly less massive progenitors at the their
  maximum luminosity. Figure~7 in \citet{sch07} shows for instance
  that PNe with core masses of 0.565~\msun\ (initial masses
  $\sim$1.5~\msun) are not able to reach the \oiii\ luminosities of
  our targets. We adopt this value as the
  lower limit for the PN progenitors studied in this work. On the
  other side, it should be considered that with increasing mass the
  number of progenitors decreases (according to the standard IMF), and
  the duration of the high \oiii\ luminosity phase of their nebulae
  becomes much shorter \citep{sch07}. Therefore, statistically, it is
  unlikely to find very massive progenitors among our targets. A rough
  upper limit of $\sim$2.5~\msun\ may be assumed, considering that none
  of them is a type~I PN \citep{p01}.
We conclude that the PNe presented in this work are expected to be
produced by stars with masses roughly between 1.5~\msun\ and
2.5~\msun. The lifetime of these stars is $\le$3.5~Gyr. This mass range
is also consistent with the modest nitrogen enhancement of the target
nebulae, because the solar-metallicity models of \citet{k10} predict
significant N overabundances only for significantly more massive
progenitors ($>$4~\msun) such as observed in some Galactic bipolar
nebulae \citep{cs95}.

It is important to note that our arguments are purely based on
single-star evolution considerations. They do not tackle the problem
that the PNLF cutoff magnitude is the same in all galaxies, even in
older stellar populations where $\ge$2~\msun\ stars are not found
\citep{c02,c10}. This may require alternative channels of PN
production to populate the bright end of the PNLF, such as mergers or
mass accretion in interacting binaries \citep{c05,s06}. In such case,
our mass estimates would not be valid. However, low-mass progenitors
would not be expected to have the high metallicity that we measure. In
addition, as discussed in the next section, there is independent
evidence that in the outer regions of M31 2~\msun\ stars with solar
metallicity do exist.

\section{Conclusions}

We have identified two PNe at very large galactocentric distances with
solar oxygen abundances and radial velocities consistent with that of
the extended disk of M31.  Their stellar progenitors, according to
single-star evolutionary theories, are estimated to be in the range
1.5--2.5~\msun. These two additional objects support our previous
conclusion (paper II) that the luminous PNe found outside the disk of
M31 trace a burst of star formation that occurred $<$3~Gyr ago.
This interpretation is also fully consistent with the results of
\citet{b15}, who detect trace of such a recent starburst in all the
HST fields studied, which span a similar galactocentric distance range
as our PNe. A similar starburst of solar metallicity stars is also
found throughout the disk of M31 by the Panchromatic Hubble Andromeda
Treasury (Ben Williams, private communication), as well as in the
outer disk of M33 \citep{b12}.  The most likely explanation for the
luminous oxygen-rich PNe in M31's exodisk is that the observed,
relatively massive stars found in the outer regions of M31 are part of
the thin disk that has been kicked out during a recent encounter with
M33, and/or following the impact with the unidentified progenitor of
the Giant Stellar Stream \citep[e.g.][]{mc09,b12,b15}.
  
Three other PNe in the outer regions of M31 have O/H abundances
$\sim$0.4 dex lower, and show significant deviations from the
kinematics of M31's exodisk. It is possible that they belong to
one or another of the halo's substructures discussed in the
literature. 
No bona fide PN belonging to the smooth, mostly metal-poor, ancient
halo described by \citet{ib14} has been found yet. Nor would this be
expected since the ancient PNe will be far fainter than the younger
ones. Assuming a total luminosity of the smooth halo component of
$\sim$10$^{-9}$~\lsun\ \citep{ib14} and a luminosity-specific PN
number of $\sim$10$^{-7}$ \citep{bu06}, some 10$^2$ halo PNe are
expected, but only a tenth of them within two magnitudes from the
bright cutoff of the PNLF,
which is the luminosity range that we have explored so
far.  The number could be even smaller considering that, in  old stellar
populations, AGB stars with core masses $\le$0.55~\msun\ may not
produce PNe, either because the post-AGB evolution of the core is too
slow to ``light up'' the nebulae before they disperse, or because
stars escape the AGB phase \citep{bu06}.

Given the complexity of the outer regions of M31 and the small number
of PNe available, our interpretation should be considered as
speculative.  However, it fits well into the modern view of a rich
interaction and merger history for M31, providing independent support
to the results obtained using other classes of stars, and adding
complementary information about chemical elements that are best
studied in ionized nebulae.

\acknowledgments

The results of this paper are based on observations made with (1) the 
Gran Telescopio Canarias (GTC), installed at the Spanish Observatorio
del Roque de los Muchachos of the Instituto de Astrof\'\i sica de
Canarias, in the island of La Palma and (2) the 3.5~m telescope at
Apache Point Observatory in Sunspot, New Mexico.  The Apache Point
Observatory 3.5-meter telescope is owned and operated by the
Astrophysical Research Consortium.  R.L.M.C. acknowledges funding from the
Spanish AYA2012-35330 grant. B.B., K.B.K., and R.B.C.H. are grateful
to our institutions and to the NSF for support under grants
AST-0806490, AST-0808201, AST-0806577, respectively.  This research
has made use of the USNOFS Image and Catalogue Archive operated by the
United States Naval Observatory, Flagstaff Station
(\anchor{http://www.nofs.navy.mil/data/fchpix/}{http://www.nofs.navy.mil/data/fchpix/}).




{\it Facilities:} \facility{GTC (OSIRIS)}; \facility{APO (DIS); \facility{SDSS SkyServer}}

\appendix
\section{Flux measurements and computed properties}

Table~\ref{T-fluxes} lists the emission-line measurements of all
observed PNe.  Column entries are as follows: the first column lists
the ion and wavelength designation of each line; f($\lambda$) gives
the value of the reddening function, normalized to H$\beta$ = 0; and
the following lines show the measured flux relative to H$\beta$ =
100. At the bottom of each column, for each nebula we list the log of
the total observed H$\beta$ flux through the spectrograph slit. Note
that the latter may differ from the magnitude in \citet{m06}, as not
all observations were obtained in photometric nights and no attempt
was done to account for slit slosses. Table~\ref{T-tene} contains the
logarithmic reddening parameter, c(H$\beta$), and the derived \Ne\ and
\Te\ of each nebula and the corresponding
diagnostics. Tables~\ref{T-ion073839}--\ref{T-ion6688372} display
their ionic abundances. They also show the temperatures adopted from
Table~\ref{T-tene} that were used to calculate them. Asterisks denote
values that were used in the brightness-weighted mean values (wm) in
lines below them. Also included for each elements is the derived
ionization correction factor ({\it icf}), calculated as described in
\citet{kh01}. Table~\ref{T-totabund} lists the total abundances.  
  Note that the ELSA code determines uncertainties in the dereddened
  line intensities including the uncertainty in
  c(H$\beta$). Uncertainties in temperature and density diagnostics
  incorporate the errors in their constituent lines, and ionic
  abundance uncertainties include errors in the relevant line
  intensities, temperature, and density. Final abundances incorporate
  errors in the ionic abundances, but not on the {\it icf}. For more
  details, see for example \citet{mi10}.  
  Note also that the calculation of a reliable sulfur abundance with 
  the adopted {\it icf} scheme requires the measurement of S$^{2+}$,
  that is estimated to be significantly more abundant than S$^+$ in
  all targets.  Therefore, in the cases where no S$^{2+}$ lines could be
  measured we do not quote the S/H and S/O total abundances.

\begin{deluxetable}{l@{}rrrrrrrrrr}
\tabletypesize{\footnotesize}
\tablecolumns{11}
\tablewidth{0pc}
\tablecaption{Line identification and observed fluxes\label{T-fluxes}}
\tablehead{
Ident.   & f($\lambda$) & \mc{1}{c}{M2507} & \mc{1}{c}{M2538} & \mc{1}{c}{M2539} & \mc{1}{c}{M2541} & \mc{1}{c}{M2543} & \mc{1}{c}{M2549} & \mc{1}{c}{M2566} & \mc{1}{c}{M2988}  & \mc{1}{c}{M31-372}
}
\startdata
\oii               \,$\lambda$ 3727 &  0.292 & 24.8     & 61.5     & 20.9       & 33.1    & 26.8     & 27.4      & 84.9        & 27.1  & \nd  \\  
\heii+H9           \,$\lambda$ 3835 &  0.262 & 4.55     & \nd      & 6.03       & \nd     & \nd      & \nd       & \nd         & \nd   & \nd  \\
\neiii             \,$\lambda$ 3869 &  0.252 & 44.2     & 107      & 84.9       & 61.9    & 93.1     & 94.9      & 108         & 68.2  & 48.3 \\
\hei+H8            \,$\lambda$ 3889 &  0.247 & 13.8     & 8.83:    & 16.7       & 15.1    & 16.0     & 17.6      & 19.5        & 13.2  & 18.8 \\
\neiii+H$\epsilon$ \,$\lambda$ 3968 &  0.225 & 28.9     & 33.0:    & 36.4       & 36.8    & 42.3     & 46.5      & 51.5        & 34.2  & 34.1 \\
\hei+\heii         \,$\lambda$ 4026 &  0.209 & \nd      & \nd      & 2.22       & \nd     & \nd      & \nd       & \nd         & \nd   & \nd  \\
\sii               \,$\lambda$ 4071 &  0.196 & \nd      & \nd      & 1.24       & \nd     & \nd      & 4.76      & 3.56        & \nd   & \nd  \\
\heii              \,$\lambda$ 4100 &  0.188 & \nd      & 0.116\ta & \nd        & 0.435\ta& \nd      & 0.308\ta  & 6.26(-2)\ta & \nd   & \nd  \\
H$\delta$          \,$\lambda$ 4101 &  0.188 & 24.3     & 29.3\ta  & \nd        & 25.2\ta & \nd      & 25.0\ta   & 25.8\ta     & \nd   & 22.6 \\
\heii              \,$\lambda$ 4339 &  0.124 & \nd      & 0.215\ta & 2.27(-2)\ta& 0.775\ta& 0.105\ta & 0.557\ta  & 0.113\ta    & \nd   & \nd  \\
H$\gamma$          \,$\lambda$ 4340 &  0.124 & 44.5     & 46.4\ta  & 46.0\ta    & 43.2\ta & 45.0\ta  & 44.2\ta   & 43.5\ta     & 48.6  & 42.2 \\
\oiii              \,$\lambda$ 4363 &  0.118 & 6.39     & 18.6     & 12.9       & 13.5    & 8.33     & 16.0      & 8.90        & 9.96  & 10.3 \\
\hei               \,$\lambda$ 4472 &  0.090 & 4.30     & 7.63::   & 4.91       & 4.21    & 5.66     & 3.79      & 5.32        & 5.35  & 6.05 \\
\heii              \,$\lambda$ 4542 &  0.072 & \nd      & \nd      & \nd        & 1.62    & \nd      & \nd       & \nd         & \nd   & \nd  \\
\rniii+\roii       \,$\lambda$ 4640 &  0.048 & \nd      & \nd      & \nd        & 1.14    & \nd      & 2.70      & \nd         & \nd   & \nd  \\
\rciii+\roii       \,$\lambda$ 4650 &  0.045 & \nd      & \nd      & \nd        & 0.901   & \nd      & \nd       & \nd         & \nd   & \nd  \\
\heii              \,$\lambda$ 4686 &  0.036 & \nd      & 12.1:    & 0.971      & 32.5    & 4.62     & 23.8      & 4.95        & \nd   & \nd  \\
\hei+\ariv         \,$\lambda$ 4711 &  0.030 & \nd      & \nd      & \nd        & 2.44    & \nd      & 4.39      & 3.69        & \nd   & \nd  \\
\ariv              \,$\lambda$ 4740 &  0.023 & \nd      & 4.41     & 2.46       & 1.69    & 5.48     & 4.53      & 3.43        & \nd   & \nd  \\
\heii              \,$\lambda$ 4859 &  0.000 & \nd      & 0.488\ta & 4.85(-2)\ta& 1.62\ta & 0.226\ta & 1.20\ta   & 0.242\ta    & \nd   & \nd  \\
H$\beta$           \,$\lambda$ 4861 &  0.000 & 100      & 100\ta   & 100\ta     & 100\ta  & 100\ta   & 100\ta    & 100\ta      & 100   & 100  \\
\hei               \,$\lambda$ 4922 & -0.021 & 1.10     & \nd      & 1.38       & 1.18    & \nd      & \nd       & \nd         & \nd   & \nd  \\
\oiii              \,$\lambda$ 4959 & -0.030 & 232      & 479      & 378        & 324     & 418      & 415       & 427         & 311   & 324  \\
\oiii              \,$\lambda$ 5007 & -0.042 & 697      & 1498     & 1140       & 978     & 1270     & 1243      & 1302        & 948   & 985  \\
\fevi              \,$\lambda$ 5147 & -0.074 & \nd      & \nd      & 0.421      & \nd     & \nd      & \nd       & \nd         & \nd   & \nd  \\ 
\heii              \,$\lambda$ 5412 & -0.134 & \nd      & \nd      & \nd        & 2.88    & \nd      & 1.79      & \nd         & \nd   & \nd  \\
\cliii             \,$\lambda$ 5538 & -0.161 & \nd      & \nd      & 0.466      & \nd     & \nd      & \nd       & \nd         & \nd   & \nd  \\
\nii               \,$\lambda$ 5755 & -0.207 & 1.24     & \nd      & 0.755      & \nd     & \nd      & \nd       & 1.30        & \nd   & \nd  \\
\hei               \,$\lambda$ 5876 & -0.231 & 16.8     & 18.8::   & 17.8       & 12.3    & 17.9     & 13.7      & 17.2        & 19.4  & 17.0 \\
\oi                \,$\lambda$ 6300 & -0.313 & 2.41     & \nd      & 2.95       & 3.41    & \nd      & \nd       & \nd         & 5.26  & \nd  \\
\heii              \,$\lambda$ 6311 & -0.315 & \nd      & \nd      & \nd        & 0.123\ta& \nd      & \nd       & \nd         & \nd   & \nd  \\
\siii              \,$\lambda$ 6312 & -0.315 & 1.67     & \nd      & 1.08\ta    & 1.07\ta & \nd      & \nd       & \nd         & \nd   & \nd  \\
\oi                \,$\lambda$ 6364 & -0.325 & 0.753    & \nd      & 1.17       & 0.854   & \nd      & \nd       & \nd         & 2.15  & \nd  \\
\nii               \,$\lambda$ 6548 & -0.358 & 9.52     & 33.6::   & 7.35       & 8.42    & 9.48     & 8.48      & 27.6        & 9.89  & 5.22 \\
\heii              \,$\lambda$ 6560 & -0.360 & \nd      & 1.79\ta  & 0.146\ta   & 4.65\ta & 0.702\ta & 3.68\ta   & 0.742\ta    & \nd   & \nd  \\
H$\alpha$          \,$\lambda$ 6563 & -0.360 & 326      & 331\ta   & 311\ta     & 295\ta  & 319\ta   & 318\ta    & 314\ta      & 357   & 334  \\
\nii               \,$\lambda$ 6584 & -0.364 & 27.9     & 80.0     & 21.4       & 24.2    & 26.6     & 26.5      & 85.4        & 27.8  & 15.5 \\
\hei               \,$\lambda$ 6678 & -0.380 & 4.34     & \nd      & 4.60       & 3.32    & 4.77     & 3.92      & 4.85        & 5.07  & 4.79 \\
\sii               \,$\lambda$ 6716 & -0.387 & 0.758    & \nd      & 0.717      & 1.69    & 2.22     & 1.99      & 10.6        & 1.03  & \nd  \\
\sii               \,$\lambda$ 6731 & -0.389 & 1.56     & \nd      & 1.12       & 2.66    & 3.00     & 2.90      & 14.9        & 1.50  & \nd  \\
\hei               \,$\lambda$ 7065 & -0.443 & 10.9     & \nd      & 11.0       & 5.29    & 7.27     & 6.25      & 6.13        & 11.6  & 8.22 \\
\ariii             \,$\lambda$ 7136 & -0.453 & 10.4     & 17.7::   & 10.8       & 5.37    & 16.6     & 10.0      & 22.1        & 9.09  & 6.38 \\
\hei               \,$\lambda$ 7281 & -0.475 & 1.24     & \nd      & \nd        & \nd     & \nd      & \nd       & \nd         & \nd   & \nd  \\
\oii               \,$\lambda$ 7324 & -0.481 & 18.9     & \nd      & 7.91       & 5.58    & 4.42     & 5.27      & 8.61        & 11.8  & 6.95 \\
\ariii             \,$\lambda$ 7751 & -0.539 & 2.77     & \nd      & 2.34       & \nd     & 3.97     & 3.66      & 4.89        & 2.04  & \nd  \\
\siii              \,$\lambda$ 9532 & -0.632 & \nd      & 148::    & \nd        & \nd     & \nd      & \nd       & \nd         & \nd   & \nd  \\[4pt]
$\log$ F(H$\beta$)\tb               &        & -15.05   & -15.05   & -15.12     & -15.23  &   -15.26 & -15.26    &      -15.26 & -15.31& -15.64\\
\enddata
\tablenotetext{a}{Deblended}
\tablenotetext{b}{In erg cm$^{-2}$ s$^{-1}$ as measured in our extracted spectra}
\end{deluxetable}

\begin{deluxetable}{lr@{}lr@{}lr@{}lr@{}lr@{}lr@{}lr@{}lr@{}lr@{}l}
\tabletypesize{\small}
\tablecolumns{11}
\tablewidth{0pc}
\tablecaption{Extinction, electron temperatures and densities\label{T-tene}}
\tablehead{
   & \mc{2}{c}{M2507} & \mc{2}{c}{M2538} & \mc{2}{c}{M2539} & \mc{2}{c}{M2541} & \mc{2}{c}{M2543} 
}
\startdata
c(H$\beta$)  & \mc{2}{c}{0.17}          & \mc{2}{c}{0.18}  & \mc{2}{c}{0.12}    & \mc{2}{c}{0.06}    & \mc{2}{c}{0.13} \\ 
\Te(\oiii)   & 11300 & $\pm$650         & 12700 & $\pm$750 & 12250 & $\pm$700   & 13150 & $\pm$800   & 10250 & $\pm$500\\ 
\Te(\nii)    & 16000 & $\pm$4450        & 10300 & \tb      & 15950 & $\pm$1750  & 10300 & \tb        & 10300 & \tb     \\ 
\Te(\oii)    &       &                  &       &          &       &            & 21950 & $\pm$18100 &       &         \\ 
\Te(\sii)    &       &                  &       &          &       &            &       &            &       &         \\ 
\Ne(\sii)    & 12200 & $^{+14750}_{-12200}$ & 10000 & \tb      &  3000 & $\pm$1600  &  3050 & $\pm$1650  &  1650 & $\pm$800\\ 
\hline\\[-5pt]
             & \mc{2}{c}{M2549} & \mc{2}{c}{M2566} & \mc{2}{c}{M2988}  & \mc{2}{c}{M31-372}      & &     \\[3pt]
\hline\\[-5pt]
c(H$\beta$)  &  \mc{2}{c}{0.15}   & \mc{2}{c}{0.12}   & \mc{2}{c}{0.29}     & \mc{2}{c}{0.19}    & & \\    
\Te(\oiii)   &  12950 & $\pm$800  & 10350 & $\pm$500  & 12250 & $\pm$700    & 11950 & $\pm$650   & & \\    
\Te(\nii)    &  10300 & \tb       & 10400 & $\pm$700  & 12350 & $\pm$350\ta & 12200 & $\pm$350\ta& &  \\   
\Te(\oii)    &        &           & 14350 & $\pm$5850 &       &             &       &            & &   \\  
\Te(\sii)    &        &           & 10250 & $\pm$3500 &       &             &       &            & &  \\   
\Ne(\sii)    &   2300 & $\pm$1150 &  1900 & $\pm$950  &  2250 & $\pm$1100   & 10000 & \tb        & &   \\  
\enddata
\tablenotetext{a}{Estimated from \Te(\oiii) according to the prescription from \citet{p92}}
\tablenotetext{b}{Default  value}
\end{deluxetable}

\begin{deluxetable}{lcr@{}lr@{}lr@{}l}
\tabletypesize{\small}
\setlength{\tabcolsep}{0.07in}
\tablecolumns{8}
\tablewidth{0in}
\tablecaption{Ionic Abundances\label{T-ion073839}}
\tablehead{
\colhead{Ion} & \colhead{T$_{\mathrm{used}}$} & \multicolumn{6}{c}{Abundance}\\
\cline{3-8}\\[-7pt]
              &   & \multicolumn{2}{c}{M2507} & \multicolumn{2}{c}{M2538} & \multicolumn{2}{c}{M2539}
}
\startdata
He$^{+}$       & [O III] & 9.85 & $\pm$1.32(-2)        & 0.103 & $\pm$0.052                   & 0.112 & $\pm$0.015\\
He$^{+2}$      &         &      &                      & 1.13  & $\pm$0.36(-2)                & 9.04 & $\pm$1.34(-4)\\
icf(He)       &         & 1.00 &                       & 1.00  &                             & 1.00 &              \\[5pt]
O$^{0}$(6300)  & [N II]  & $^{*}$1.26 & $\pm$0.75(-6)  &           &                          & $^{*}$1.48 & $\pm$0.42(-6)\\     
O$^{0}$(6363)  & [N II]  & $^{*}$1.23 & $\pm$0.73(-6)  &           &                          & $^{*}$1.82 & $\pm$0.53(-6)\\     
O$^{0}$        & wm      & 1.25 & $\pm$0.74(-6)        &           &                         & 1.57 & $\pm$0.44(-6)\\                     
O$^{+}$(3727)  & [N II]  & $^{*}$7.56 & $\pm$9.57(-6)  & $^{*}$4.79 & $\pm$1.01(-5)           & $^{*}$3.06 & $\pm$1.01(-6)\\     
O$^{+}$(7325)  & [N II]  & $^{*}$1.29 & $\pm$0.78(-5)  &           &                          & $^{*}$7.96 & $\pm$3.27(-6)\\     
O$^{+}$        & wm      & 9.88 & $\pm$8.30(-6)        &           &                         & 4.41 & $\pm$1.25(-6)\\                     
O$^{+2}$(5007) & [O III] & $^{*}$1.65 & $\pm$0.43(-4)  & $^{*}$2.47 & $\pm$0.64(-4)           & $^{*}$2.08 & $\pm$0.53(-4)\\   
O$^{+2}$(4959) & [O III] & $^{*}$1.59 & $\pm$0.34(-4)  & $^{*}$2.29 & $\pm$0.48(-4)           & $^{*}$2.00 & $\pm$0.41(-4)\\   
O$^{+2}$(4363) & [O III] & $^{*}$1.65 & $\pm$0.43(-4)  & $^{*}$2.47 & $\pm$0.64(-4)           & $^{*}$2.08 & $\pm$0.53(-4)\\   
O$^{+2}$       & wm      & 1.63 & $\pm$0.40(-4)        & 2.43 & $\pm$0.59(-4)                 & 2.06 & $\pm$0.49(-4)\\                    
icf(O)        &         & 1.00 &                       & 1.11 & $\pm$0.07                    & 1.01 & $\pm$0.00\\[5pt] 
Ar$^{+2}$(7135)& [O III] & $^{*}$6.20 & $\pm$1.44(-7)  & $^{*}$8.14 & $\pm$4.42(-7)            & $^{*}$5.67 & $\pm$1.34(-7)\\  
Ar$^{+2}$(7751)& [O III] & $^{*}$6.58 & $\pm$1.69(-7)  &           &                           & $^{*}$4.98 & $\pm$1.30(-7)\\  
Ar$^{+2}$      & wm      & 6.28 & $\pm$1.44(-7)        &           &                          & 5.55 & $\pm$1.30(-7)\\                   
Ar$^{+3}$(4740)&         &      &                      & $^{*}$3.91 & $\pm$0.72(-7)           & $^{*}$2.72 & $\pm$0.52(-7)\\  
icf(Ar)       &         & 1.06 & $\pm$0.05             & 1.29 & $\pm$0.09                    & 1.03 & $\pm$0.01\\[5pt]                           
Cl$^{+2}$(5537)&         &      &                      &           &                          & $^{*}$3.29 & $\pm$0.67(-8)\\  
icf(Cl)       &         &      &                       &           &                         & 1.01 & $\pm$0.00\\[5pt]                           
N$^{+}$(6584)  & [N II]  & $^{*}$2.60 & $\pm$1.23(-6)  & $^{*}$1.16 & $\pm$0.20(-5)           & $^{*}$1.76 & $\pm$0.44(-6)\\     
N$^{+}$(6548)  & [N II]  & $^{*}$2.61 & $\pm$1.20(-6)  & $^{*}$1.44 & $\pm$0.74(-5)           & $^{*}$1.78 & $\pm$0.39(-6)\\     
N$^{+}$(5755)  & [N II]  & $^{*}$2.60 & $\pm$1.23(-6)  &           &                          & $^{*}$1.76 & $\pm$0.44(-6)\\     
N$^{+}$        & wm      & 2.61 & $\pm$1.21(-6)        & 1.24 & $\pm$0.30(-5)                & 1.76 & $\pm$0.41(-6)\\                     
icf(N)        &         & 17.5 & $\pm$12.83            & 6.72 & $\pm$1.48                   & 48.2 & $\pm$14.76\\[5pt]                     
Ne$^{+2}$(3869)& [O III] & $^{*}$3.11 & $\pm$0.77(-5)  & $^{*}$5.01 & $\pm$1.22(-5)           & $^{*}$4.33 & $\pm$1.04(-5)\\  
Ne$^{+2}$(3967)& [O III] & 6.68 & $\pm$1.62(-5)        & 5.05 & $\pm$1.88(-5)               & 6.12 & $\pm$1.45(-5)\\        
icf(Ne)       &         & 1.05 & $\pm$0.06             & 1.33 & $\pm$0.11                   & 1.02 & $\pm$0.01\\[5pt]                      
S$^{+}$        & [N II]  & $^{*}$7.32 & $\pm$7.97(-8)  &           &                         & $^{*}$2.73 & $\pm$0.84(-8)\\           
S$^{+}$(6716)  & [N II]  & 7.31 & $\pm$8.04(-8)        &           &                         & 2.70 & $\pm$0.84(-8)\\           
S$^{+}$(6731)  & [N II]  & 7.33 & $\pm$7.93(-8)        &           &                         & 2.75 & $\pm$0.84(-8)\\           
S$^{+}$        & [S II]  &      &                      &            &                        &        &              \\        
S$^{+2}$(6312) & [O III] & $^{*}$2.17 & $\pm$0.59(-6)   &           &                         & $^{*}$1.11 & $\pm$0.31(-6)              \\                                                      
S$^{+2}$(9532) & [O III] &           &                 & $^{*}$3.97 & $\pm$2.23(-6)          &      &              \\
icf(S)        &         & 1.57 & $\pm$0.48             & 1.24 & $\pm$0.04                   & 2.37 & $\pm$0.42                              
\enddata
\end{deluxetable}

\begin{deluxetable}{lcr@{}lr@{}lr@{}l}
\tabletypesize{\small}
\setlength{\tabcolsep}{0.07in}
\tablecolumns{8}
\tablewidth{0in}
\tablecaption{Ionic Abundances\label{T-ion414349}}
\tablehead{
\colhead{Ion} & \colhead{T$_{\mathrm{used}}$} & \multicolumn{6}{c}{Abundance}\\
\cline{3-8}\\[-7pt]
              &   & \multicolumn{2}{c}{M2541} & \multicolumn{2}{c}{M2543} & \multicolumn{2}{c}{M2549}
}
\startdata
He$^{+}$       & [O III] & 7.63 & $\pm$1.09(-2)       &    0.120 & $\pm$0.015        &  8.42 & $\pm$1.18(-2) \\           
He$^{+2}$      &         & 2.98 & $\pm$0.45(-2)       &    4.27 & $\pm$0.65(-3)      &  2.21 & $\pm$0.33(-2) \\           
icf(He)       &         &  1.00                     & &    1.00 &                    &  1.00 & \\[5pt]                          
O$^{0}$(6300)  & [N II]  & $^{*}$4.52 & $\pm$0.74(-6) &         &                    &      &                  \\         
O$^{0}$(6363)  & [N II]  & $^{*}$3.53 & $\pm$0.58(-6) &         &                    &      &                  \\         
O$^{0}$        & wm      & 4.32 & $\pm$0.68(-6)       &         &                    &      &                  \\         
O$^{+}$(3727)  & [N II]  & $^{*}$1.38 & $\pm$0.37(-5) &    $^{*}$1.29 & $\pm$0.30(-5) & $^{*}$1.13 & $\pm$0.28(-5) \\     
O$^{+}$(7325)  & [N II]  & $^{*}$2.93 & $\pm$0.80(-5) &    $^{*}$3.37 & $\pm$0.83(-5) & $^{*}$2.81 & $\pm$0.75(-5) \\     
O$^{+}$        & wm      & 1.60 & $\pm$0.26(-5)      &     1.58 & $\pm$0.21(-5) &       1.40 & $\pm$0.19(-5)      \\      
O$^{+2}$(5007) & [O III] & $^{*}$1.45 & $\pm$0.39(-4) &    $^{*}$4.13 & $\pm$0.99(-4) & $^{*}$1.91 & $\pm$0.50(-4) \\     
O$^{+2}$(4959) & [O III] & $^{*}$1.40 & $\pm$0.30(-4) &    $^{*}$3.94 & $\pm$0.76(-4) & $^{*}$1.85 & $\pm$0.39(-4) \\     
O$^{+2}$(4363) & [O III] & $^{*}$1.45 & $\pm$0.39(-4) &    $^{*}$4.13 & $\pm$0.99(-4) & $^{*}$1.91 & $\pm$0.50(-4) \\     
O$^{+2}$       & wm      & 1.44 & $\pm$0.35(-4)      &     4.09 & $\pm$0.91(-4) &       1.89 & $\pm$0.46(-4)      \\      
icf(O)        &         &  1.39 & $\pm$0.08          &     1.03 & $\pm$0.01 &           1.26 & $\pm$0.05          \\[5pt]       
Ar$^{+2}$(7135)& [O III] & $^{*}$2.59 & $\pm$0.62(-7) &    $^{*}$1.29 & $\pm$0.29(-6) & $^{*}$4.55 & $\pm$1.09(-7) \\     
Ar$^{+2}$(7751)& [O III] &         &                 &     $^{*}$1.24 & $\pm$0.31(-6) & $^{*}$6.68 & $\pm$1.77(-7) \\     
Ar$^{+2}$      & wm      &         &                 &     1.28 & $\pm$0.28(-6) &       5.12 & $\pm$1.22(-7)      \\      
Ar$^{+3}$(4740)&         & $^{*}$1.56 & $\pm$0.31(-7) &    $^{*}$1.03 & $\pm$0.19(-6) & $^{*}$4.47 & $\pm$0.86(-7) \\     
icf(Ar)       &         &  1.50 & $\pm$0.09          &     1.07 & $\pm$0.01 &           1.34 & $\pm$0.06          \\[5pt]             
Cl$^{+2}$(5537)&         &         &                 &          &                    &      &                  \\         
icf(Cl)       &         &          &                 &          &                    &      &                  \\[5pt]                
N$^{+}$(6584)  & [N II]  & $^{*}$3.45 & $\pm$0.60(-6) &    $^{*}$4.49 & $\pm$0.71(-6) & $^{*}$3.53 & $\pm$0.61(-6) \\     
N$^{+}$(6548)  & [N II]  & $^{*}$3.52 & $\pm$0.61(-6) &    $^{*}$4.72 & $\pm$0.73(-6) & $^{*}$3.33 & $\pm$0.57(-6) \\     
N$^{+}$(5755)  & [N II]  &         &                 &          &                    &      &                  \\         
N$^{+}$        & wm      & 3.47 & $\pm$0.57(-6)       &    4.55 & $\pm$0.65(-6) &       3.48 & $\pm$0.57(-6)       \\     
icf(N)        &         &  13.9 & $\pm$2.53           &    27.8 & $\pm$5.59 &           18.3 & $\pm$3.28 \\[5pt]               
Ne$^{+2}$(3869)& [O III] & $^{*}$2.39 & $\pm$0.59(-5) &    $^{*}$9.43 & $\pm$2.18(-5) & $^{*}$4.04 & $\pm$0.99(-5) \\      
Ne$^{+2}$(3967)& [O III] & 4.70 & $\pm$1.14(-5)      &     1.41 & $\pm$0.32(-4) &       6.51 & $\pm$1.57(-5) \\           
icf(Ne)       &         &  1.52 & $\pm$0.11          &     1.07 & $\pm$0.01 &           1.34 & $\pm$0.07 \\[5pt]                
S$^{+}$        & [N II]  & $^{*}$1.09 & $\pm$0.28(-7) &    $^{*}$1.32 & $\pm$0.25(-7) & $^{*}$1.05 & $\pm$0.24(-7) \\     
S$^{+}$(6716)  & [N II]  & 1.10 & $\pm$0.28(-7)      &     1.32 & $\pm$0.25(-7)      &  1.06 & $\pm$0.24(-7) \\                
S$^{+}$(6731)  & [N II]  & 1.08 & $\pm$0.28(-7)      &     1.32 & $\pm$0.25(-7)      &  1.04 & $\pm$0.24(-7) \\                
S$^{+}$        & [S II]  &      &               &            &                     &        &              \\        
S$^{+2}$(6312) & [O III] & $^{*}$8.88 & $\pm$2.65(-7) &         &                    &      &                  \\                                       
S$^{+2}$(9532) & [O III] &         &                 &          &                    &      &                  \\         
icf(S)        &         &  1.44 & $\pm$0.08           &    1.77 & $\pm$0.17         &   1.57 & $\pm$0.11   
\enddata
\end{deluxetable}

\begin{deluxetable}{lcr@{}lr@{}lr@{}l}
\tabletypesize{\small}
\setlength{\tabcolsep}{0.07in}
\tablecolumns{8}
\tablewidth{0in}
\tablecaption{Ionic Abundances\label{T-ion6688372}}
\tablehead{
\colhead{Ion} & \colhead{T$_{\mathrm{used}}$} & \multicolumn{6}{c}{Abundance}\\
\cline{3-8}\\[-7pt]
              &   & \multicolumn{2}{c}{M2566} & \multicolumn{2}{c}{M2988} & \multicolumn{2}{c}{M31-372}
}
\startdata
He$^{+}$       & [O III] & 0.116 & $\pm$0.015        &   0.115 & $\pm$0.015        &  9.73 & $\pm$1.28(-2) \\        
He$^{+2}$      &         &  4.57 & $\pm$0.69(-3) &           &                     &        &              \\        
icf(He)       &         &  1.00 & &                      1.00 & &                     1.00 &             \\[5pt]                       
O$^{0}$(6300)  & [N II]  &     &                     &   $^{*}$4.05 & $\pm$0.82(-6) &       &              \\        
O$^{0}$(6363)  & [N II]  &     &                     &   $^{*}$5.13 & $\pm$1.06(-6) &       &              \\        
O$^{0}$        & wm      &     &                     &   4.36 & $\pm$0.84(-6) &             &              \\           
O$^{+}$(3727)  & [N II]  & $^{*}$3.98 & $\pm$1.20(-5) &  $^{*}$7.50 & $\pm$1.89(-6) &       &              \\        
O$^{+}$(7325)  & [N II]  & $^{*}$6.03 & $\pm$2.59(-5) &  $^{*}$2.58 & $\pm$0.85(-5) & $^{*}$1.16 & $\pm$0.32(-5) \\  
O$^{+}$        & wm      & 4.17 & $\pm$1.16(-5) &        1.30 & $\pm$0.27(-5) &             &              \\           
O$^{+2}$(5007) & [O III] & $^{*}$4.10 & $\pm$0.99(-4) &  $^{*}$1.72 & $\pm$0.44(-4) & $^{*}$1.96 & $\pm$0.50(-4) \\  
O$^{+2}$(4959) & [O III] & $^{*}$3.90 & $\pm$0.75(-4) &  $^{*}$1.64 & $\pm$0.34(-4) & $^{*}$1.88 & $\pm$0.38(-4) \\  
O$^{+2}$(4363) & [O III] & $^{*}$4.10 & $\pm$0.99(-4) &  $^{*}$1.72 & $\pm$0.44(-4) & $^{*}$1.96 & $\pm$0.50(-4) \\  
O$^{+2}$       & wm      & 4.05 & $\pm$0.90(-4) &        1.70 & $\pm$0.41(-4) &       1.94 & $\pm$0.46(-4) \\           
icf(O)        &         &  1.04 & $\pm$0.01 &            1.00 & &                     1.00 &                \\[5pt]                       
Ar$^{+2}$(7135)& [O III] & $^{*}$1.70 & $\pm$0.38(-6) &  $^{*}$4.05 & $\pm$0.96(-7) & $^{*}$3.34 & $\pm$0.77(-7) \\  
Ar$^{+2}$(7751)& [O III] & $^{*}$1.52 & $\pm$0.38(-6) &  $^{*}$3.56 & $\pm$0.93(-7) &       &              \\        
Ar$^{+2}$      & wm      & 1.67 & $\pm$0.37(-6) &        3.96 & $\pm$0.93(-7) &             &              \\          
Ar$^{+3}$(4740)&         & $^{*}$6.22 & $\pm$1.13(-7) &      &                     &        &              \\        
icf(Ar)       &         &  1.14 & $\pm$0.04 &            1.08 & $\pm$0.01 &           1.06 & $\pm$0.01          \\[5pt]                       
Cl$^{+2}$(5537)&         &     &                     &       &                     &        &              \\        
icf(Cl)       &         &      &                     &       &                     &        &                  \\[5pt]                   
N$^{+}$(6584)  & [N II]  & $^{*}$1.43 & $\pm$0.35(-5) &  $^{*}$2.65 & $\pm$0.52(-6) & $^{*}$1.84 & $\pm$0.35(-6) \\  
N$^{+}$(6548)  & [N II]  & $^{*}$1.36 & $\pm$0.29(-5) &  $^{*}$2.79 & $\pm$0.54(-6) & $^{*}$1.82 & $\pm$0.34(-6) \\  
N$^{+}$(5755)  & [N II]  & $^{*}$1.43 & $\pm$0.35(-5) &      &                     &        &              \\        
N$^{+}$        & wm      & 1.41 & $\pm$0.33(-5) &        2.69 & $\pm$0.49(-6) &       1.83 & $\pm$0.33(-6) \\           
icf(N)        &         &  11.1 & $\pm$3.23 &            14.0 & $\pm$1.87 &           17.8 & $\pm$3.62         \\[5pt]            
Ne$^{+2}$(3869)& [O III] & $^{*}$1.04 & $\pm$0.24(-4) &  $^{*}$3.86 & $\pm$0.93(-5) & $^{*}$2.82 & $\pm$0.68(-5) \\  
Ne$^{+2}$(3967)& [O III] & 1.64 & $\pm$0.37(-4) &        6.33 & $\pm$1.50(-5) &       6.55 & $\pm$1.54(-5)     \\    
icf(Ne)       &         &  1.14 & $\pm$0.04 &            1.04 & $\pm$0.01 &           1.00 &                    \\[5pt]         
S$^{+}$        & [N II]  & $^{*}$6.58 & $\pm$1.71(-7) &  $^{*}$3.91 & $\pm$0.94(-8) &       &              \\        
S$^{+}$(6716)  & [N II]  & 6.58 & $\pm$1.72(-7) &        3.91 & $\pm$0.94(-8) &             &              \\        
S$^{+}$(6731)  & [N II]  & 6.58 & $\pm$1.70(-7) &        3.91 & $\pm$0.95(-8) &             &              \\        
S$^{+}$        & [S II]  & 6.77 & $\pm$5.06(-7) &            &                     &        &              \\        
S$^{+2}$(6312) & [O III] &     &                     &       &                     &        &              \\                                             
S$^{+2}$(9532) & [O III] &     &                     &       &                     &        &              \\        
icf(S)        &         &  1.35 & $\pm$0.08          &   1.60 & $\pm$0.11         &         &               
\enddata
\end{deluxetable}

\begin{deluxetable}{lr@{}lr@{}lr@{}lr@{}lr@{}l}
\tabletypesize{\small}
\setlength{\tabcolsep}{0.07in}
\tablecolumns{5}
\tablewidth{0pc}
\tablecaption{Total Elemental Abundances\label{T-totabund}}
\tablehead{
\colhead{Element} &
\multicolumn{2}{c}{M2507} &
\multicolumn{2}{c}{M2538} &
\multicolumn{2}{c}{M2539} &
\multicolumn{2}{c}{M2541} &
\multicolumn{2}{c}{M2543} 
}
\startdata
He/H & 9.85 & $\pm$1.32(-2) & 0.115& $\pm$0.053    & 0.113& $\pm$0.015    & 0.106& $\pm$0.012    & 0.125& $\pm$0.015    \\ 
N/H  & 4.56 & $\pm$1.67(-5) & 8.36 & $\pm$3.26(-5) & 8.50 & $\pm$2.34(-5) & 4.81 & $\pm$1.43(-5) & 1.26 & $\pm$0.37(-4) \\ 
N/O  & 0.264& $\pm$0.107    & 0.259& $\pm$0.089    & 0.400& $\pm$0.056    & 0.216& $\pm$0.047    & 0.288& $\pm$0.057    \\ 
O/H  & 1.73 & $\pm$0.44(-4) & 3.22 & $\pm$0.71(-4) & 2.12 & $\pm$0.50(-4) & 2.22 & $\pm$0.51(-4) & 4.39 & $\pm$0.95(-4) \\ 
Ne/H & 3.26 & $\pm$0.85(-5) & 6.64 & $\pm$1.63(-5) & 4.43 & $\pm$1.06(-5) & 3.64 & $\pm$0.93(-5) & 1.01 & $\pm$0.23(-4) \\ 
Ne/O & 0.188& $\pm$0.031    & 0.206& $\pm$0.033    & 0.209& $\pm$0.034    & 0.164& $\pm$0.027    & 0.229& $\pm$0.036    \\ 
S/H  & 3.52 & $\pm$1.37(-6) & 4.93 & $\pm$2.83(-6) & 2.69 & $\pm$1.05(-6) & 1.44 & $\pm$0.46(-6) & \nd  & \nd           \\ 
S/O  & 2.04 & $\pm$0.81(-2) & 1.53 & $\pm$0.88(-2) & 1.27 & $\pm$0.33(-2) & 6.46 & $\pm$1.58(-3) & \nd  & \nd           \\ 
Cl/H & \nd  & \nd           & \nd  & \nd           & 3.32 & $\pm$0.67(-8) & \nd  & \nd           & \nd  & \nd           \\ 
Cl/O & \nd  & \nd           & \nd  & \nd           & 1.56 & $\pm$0.23(-4) & \nd  & \nd           & \nd  & \nd           \\ 
Ar/H & 6.66 & $\pm$1.60(-7) & 1.55 & $\pm$0.58(-6) & 8.51 & $\pm$1.61(-7) & 6.22 & $\pm$1.05(-7) & 2.49 & $\pm$0.40(-6) \\ 
Ar/O & 3.85 & $\pm$0.81(-3) & 4.81 & $\pm$1.78(-3) & 4.01 & $\pm$0.63(-3) & 2.80 & $\pm$0.43(-3) & 5.66 & $\pm$0.80(-3) \\ 
\hline \\
\colhead{       } &
\multicolumn{2}{c}{M2549} &
\multicolumn{2}{c}{M2566} &
\multicolumn{2}{c}{M2988} &
\multicolumn{2}{c}{M31-372} &
\multicolumn{1}{c}{\it Solar ref.} &
\multicolumn{1}{c}{\it Orion ref.}\\[3pt]
\hline\\[-5pt]
He/H & 0.106& $\pm$0.013    & 0.121& $\pm$0.015    & 0.115& $\pm$0.015    & 9.73 & $\pm$1.28(-2) & 8.50(-2)\ph{x.} & \ph{xx}9.70(-2)\\
N/H  & 6.36 & $\pm$1.89(-5) & 1.57 & $\pm$0.47(-4) & 3.77 & $\pm$0.88(-5) & 3.26 & $\pm$0.70(-5) & 6.76(-5)\ph{x.} & \ph{xx}5.37(-5)\\
N/O  & 0.248& $\pm$0.048    & 0.339& $\pm$0.077    & 0.206& $\pm$0.032    & 0.158& $\pm$0.026    & 0.138   \ph{xx} & \ph{xx}0.100   \\
O/H  & 2.57 & $\pm$0.59(-4) & 4.64 & $\pm$0.97(-4) & 1.83 & $\pm$0.43(-4) & 2.06 & $\pm$0.48(-4) & 4.89(-4)\ph{x.} & \ph{xx}5.37(-4)\\
Ne/H & 5.40 & $\pm$1.35(-5) & 1.19 & $\pm$0.27(-4) & 4.03 & $\pm$0.97(-5) & 2.82 & $\pm$0.68(-5) & 8.51(-5)\ph{x.} & \ph{xx}1.12(-4)\\
Ne/O & 0.210& $\pm$0.034    & 0.255& $\pm$0.041    & 0.220& $\pm$0.036    & 0.137& $\pm$0.022    & 0.174   \ph{xx} & \ph{xx} 0.209  \\
S/H  & \nd  & \nd           & \nd  & \nd           & \nd  & \nd           & \nd  & \nd           & 1.32(-5)\ph{x.} & \ph{xx}1.66(-5)\\
S/O  & \nd  & \nd           & \nd  & \nd           & \nd  & \nd           & \nd  & \nd           & 2.70(-2)\ph{x.} & \ph{xx}3.09(-2)\\
Cl/H & \nd  & \nd           & \nd  & \nd           & \nd  & \nd           & \nd  & \nd           & 3.16(-7)\ph{x.} & \ph{xx}2.88(-7)\\                               
Cl/O & \nd  & \nd           & \nd  & \nd           & \nd  & \nd           & \nd  & \nd           & 6.46(-4)\ph{x.} & \ph{xx}5.36(-4)\\                         
Ar/H & 1.28 & $\pm$0.21(-6) & 2.62 & $\pm$0.45(-6) & 4.26 & $\pm$1.00(-7) & 3.54 & $\pm$0.83(-7) & 2.51(-6)\ph{x.} & \ph{xx}4.17(-6)\\
Ar/O & 4.99 & $\pm$0.70(-3) & 5.64 & $\pm$0.92(-3) & 2.33 & $\pm$0.48(-3) & 1.72 & $\pm$0.36(-3) & 5.13(-3)\ph{x.} & \ph{xx}7.77(-3)\\
\enddata
\end{deluxetable}

\end{document}